\documentstyle[aps,epsfig,preprint]{revtex}
\begin{document}
\tolerance=10000
\hbadness=10000
\tighten
\draft
\title{Dipole resonances in oxygen isotopes in time-dependent
density-matrix 
theory
}
\author{M. Tohyama}
\address{Kyorin University School of Medicine, 
Mitaka, Tokyo 181-8611, Japan}
\author{A. S. Umar}
\address{Department of Physics and Astronomy, Vanderbilt University,
Nashville, Tennessee 37235, USA}
\date{\today}
\maketitle
\begin{abstract}
The strength functions of isovector dipole resonances
in the even oxygen isotopes $^{18\sim24}$O are calculated by
using an extended version of the time-dependent Hartree-Fock theory
known as the time-dependent 
density-matrix theory (TDDM). The results are compared with recent
experimental data and also with a shell-model calculation. It is found
that the observed isotope dependence of low-lying dipole strength is reproduced
in TDDM when the strength of the residual interaction is appropriately chosen.
It is also found that there is a difference between the TDDM prediction
and the shell-model calculation for $^{24}$O.

\vspace{0.5cm}

\vspace{0.5cm}
\end{abstract}
\newpage
The study of structures and reactions of nuclei far from stability has been
one of the most active fields of nuclear physics in the past decade \cite{RIA}.
The microscopic description of such nuclei will lead to a better understanding
of the interplay among the strong, Coulomb, and the weak interactions as well
as the enhanced correlations present in these many-body systems.
For neutron rich nuclei, in the extreme limit, the proton and neutron
densities may exhibit strikingly different behavior.
One of the observables for studying this phenomenon is the evolution of the
isovector dipole modes along an isotope chain.
Several years ago explorative calculations were performed for isovector dipole
($E1$) and isoscalar quadrupole ($E2$) resonances
in the unstable oxygen isotope $^{22}$O
based on the time-dependent density-matrix theory (TDDM) \cite{Toh1}.
It was found that there are low-energy $E1$ and $E2$ modes
associated with excess neutrons.
At that time there were no experimental data for comparison
and the TDDM calculations were not highly quantitative since spin-orbit 
force was neglected in the calculation of a mean-field potential,
and the neutron 2$s$ and 1$d$ states were assumed to be fractionally occupied, 
apparently overestimating the effect of ground-state correlations. 
Recently the isotope dependence of low-lying $E1$ strength
has been  measured at GSI \cite{Dat} for oxygen isotopes 
and compared with a shell-model
calculation \cite{Sag}.
The aim of this paper is to present more quantitative
calculations of $E1$ modes for these isotopes using the time-dependent
Hartree-Fock (TDHF) and the TDDM theories.
Two recent improvements make it
possible to perform better TDDM calculations: one is the treatment of
ground-state correlations. As will be described below, an
adiabatic treatment of the residual interaction enables us to
obtain
a correlated ground state which is a stationary solution of TDDM \cite{Toh2}.
The other improvement is the extension of 
the TDDM program to include
spin-orbit force, as it was done for TDHF calculations \cite{Uma}.

TDDM is an extended version of TDHF
and is formulated in order to determine the time evolution of 
one-body and two-body 
density matrices $\rho$ and $\rho_2$ in 
a self-consistent manner \cite{Gon}.
TDDM, therefore, includes the effects of both a mean-field potential 
and two-body correlations.
The equations of motion for $\rho$ and $\rho_2$ can be derived
by truncating the well-known BBGKY hierarchy for
reduced density matrices \cite{Wan}.
To solve the equations of motion for $\rho$ and $\rho_2$ ,
we expand $\rho$ and $C_2$ (the correlated part of $\rho_2$)
using a finite number of single-particle states $\psi_{\alpha}$
which satisfy a TDHF-like equation,
\begin{eqnarray}
\rho(11',t)=\sum_{\alpha\alpha'}n_{\alpha\alpha'}(t)\psi_{\alpha}(1,t)
\psi_{\alpha'}^{*}(1',t), 
\end{eqnarray}
\begin{eqnarray}
C_{2}(121'2',t)&=&\rho_{2} - A(\rho\rho) \nonumber \\
&=&\sum_{\alpha\beta\alpha'\beta'}C_{\alpha\beta\alpha'\beta'}(t)
\nonumber \\
&\times&\psi_{\alpha}(1,t)\psi_{\beta}(2,t)
\psi_{\alpha'}^{*}(1',t)\psi_{\beta'}^{*}(2',t), 
\end{eqnarray}
where the numbers denote space, spin and isospin coordinates.
Thus, the equations of motion of TDDM consist of the following three
coupled equations \cite{Gon}:
\begin{eqnarray}
i\hbar\frac{\partial}{\partial t}\psi_{\alpha}(1,t)=h(1,t)
\psi_{\alpha}(1,t),
\end{eqnarray}
\begin{eqnarray}
i\hbar \dot{n}_{\alpha\alpha'}=\sum_{\beta\gamma\delta}
[\langle\alpha\beta|v|\gamma\delta\rangle C_{\gamma\delta\alpha'\beta}
-C_{\alpha\beta\gamma\delta}\langle\gamma\delta|v|\alpha'\beta\rangle],
\end{eqnarray}
\begin{eqnarray}
i\hbar\dot{C}_{\alpha\beta\alpha'\beta'}=B_{\alpha\beta\alpha'\beta'}
+P_{\alpha\beta\alpha'\beta'}+H_{\alpha\beta\alpha'\beta'}, 
\end{eqnarray}
where $h$ is the mean-field Hamiltonian and $v$ the residual interaction.
The terms on 
the right-hand side of Eq.(5) contain all the two-body correlations including
those induced by the Pauli exclusion principle \cite{Toh1}.
The TDDM equations of motion satisfy conservation laws of
the total number of particles and the total momentum and energy.
The small amplitude limit of TDDM was investigated \cite{Toh3} and
it was found that if only the 1 particle - 1 hole and 1 hole - 1 particle
elements of $n_{\alpha\alpha'}$ and the 2 particle - 
2 hole and 2 hole - 2 particle elements of $C_{\alpha\beta\alpha'\beta'}$
are taken,
the small amplitude limit of 
TDDM is equivalent to
the conventional second RPA (SRPA) \cite{Saw}.
Thus, TDDM is a more general framework than the conventional SRPA.

The $E1$ strength function is calculated 
according to the following three
steps:

1) A static Hartree-Fock (HF) calculation is performed to obtain
the initial ground state. The Skyrme III with spin-orbit force
is used as the effective interaction. 
It has been reported \cite{Otsu} that the Skyrme III is useful
even for very neutron rich nuclei.
Unoccupied single-particle
states up to the $2s$ and $1d$ orbits are also calculated 
for both protons and neutrons to solve
the TDDM equations for
$n_{\alpha\alpha'}$ and $C_{\alpha\beta\alpha'\beta'}$. 
The single-particle wavefunctions are confined to a cylinder
with length 20fm and radius 10fm. (Axial symmetry is imposed to
calculate the single-particle wavefunctions \cite{Uma}.)
The mesh size used is 0.5fm.
The neutron $1d_{5/2}$ state is assumed to be partially
occupied to obtain the HF ground states for $^{18}$O and $^{20}$O.

2) To obtain a correlated ground state, we evolve the HF ground state 
using the TDDM equations and the following time-dependent residual
interaction of the $\delta$-function form
\begin{eqnarray}
v(t)=v_0(1-e^{-t/\tau})\delta^3(\vec{r}-\vec{r'}).
\end{eqnarray}
The time constant $\tau$ should be sufficiently large 
to obtain a nearly stationary solution of the TDDM equations \cite{Toh2}.
We choose $\tau$ to be 150fm/c. The strength of the residual interaction 
is determined
to approximately reproduce the observed occupation probability of 
the proton $1d_{5/2}$ state
in $^{16}$O \cite{Leu}. The obtained value of $v_0$ 
is $-230$MeVfm$^{3}$, which might be smaller than the value of 
approximately $-300$MeVfm$^3$ 
found in the literature for
pairing-gap calculations \cite{Toh4}. 
The time step size used to solve the TDDM equations is
0.75fm/c.
The calculated occupation probabilities of the neutron 
$2s_{1/2}$ and $1d_{3/2}$ are $2\sim3\%$
in the oxygen isotopes considered here.

3) The $E1$ mode is excited by boosting the single-particle 
wavefunctions at $t=5\tau$ with the dipole velocity field:
\begin{eqnarray}
\psi_{\alpha}(5\tau)\longrightarrow e^{ikD(z)}\psi_{\alpha},
\end{eqnarray}
where
\begin{eqnarray}
D(z)=
 \frac{N}{A}z,~~ (- \frac{Z}{A}z )
~~ \mbox{for protons}~~ (\mbox{for neutrons}).
%\begin{array}{ll}
% \frac{N}{A}z & \mbox{for protons} \\
% - \frac{Z}{A}z & \mbox{for neutrons.}
%\end{array}
%\right.
\end{eqnarray}
Here, $Z$ and $N$ are the numbers of protons and neutrons, respectively, and
$A=Z+N$. 
When the boosting parameter $k$ is sufficiently small,
the strength function defined by
\begin{eqnarray}
S(E)=\sum_{n}|\langle\Phi_n|\hat{D}|\Phi_0\rangle|^{2}\delta (E-E_{n})
\end{eqnarray}
is obtained from the Fourier transformation of the time-dependent
dipole moment $D(t)$ as
\begin{eqnarray}
S(E)=\frac{1}{\pi k\hbar}\int_{0}^{\infty}D(t)\sin\frac{Et}{\hbar}
dt, 
\end{eqnarray}
where
\begin{eqnarray}
D(t)
=\int D(z)\rho(\vec{r},t)d^3\vec{r}.
\end{eqnarray}
In Eq.(9) $|\Phi_0\rangle$ is the total ground-state wavefunction and
$|\Phi_n\rangle$ the wavefunction for 
an excited state with excitation energy
$E_n$.
It is very time consuming to solve
the TDDM equations (especially Eq.(5)) for a long period.
Thus, we stop TDDM calculations at
$t=1200$fm/c. The upper limit of
the time integration in Eq.(10) is limited to 450fm/c.
To reduce fluctuations in $S(E)$, 
the dipole moment is multiplied by a damping factor 
$e^{-\Gamma t/2\hbar}$
with $\Gamma=1$MeV before the time integration.
Since the integration time is limited, the strength function in a
very low energy region ($E<2\pi\hbar/450\approx3$MeV) is not well
determined.
The energy-weighted sum rule (EWSR) is expressed as
\begin{eqnarray}
\int S(E)EdE
&=& \frac{1}{2}\langle\Phi_0|[\hat{D},[H,\hat{D}]]|\Phi_0\rangle
\nonumber \\
&=& \frac{\hbar^2}{2m}\frac{ZN}{A} \nonumber \\
+
\frac{t_1+t_2}{4}(\int\rho_p(\vec{r})\rho_n(\vec{r})d^3\vec{r}
&+&\sum_{\alpha\alpha'\in p, \beta\beta'\in n}
\int\psi_{\alpha}^*(\vec{r})\psi_{\beta}^*(\vec{r})
\psi_{\alpha'}(\vec{r})\psi_{\beta'}(\vec{r})d^3\vec{r}
C_{\alpha'\beta'\alpha\beta}),
\end{eqnarray}
where $m$ is the nucleon mass, and $t_1$ and $t_2$ are the parameters 
for the momentum dependent parts of the Skyrme interaction.
We assume that the Hamiltonian $H$ consists of a two-body interaction
of the Skyrme type.
The first term on the right-hand side of the above equation corresponds
to the classical Thomas-Reiche-Kuhn (TRK) sum rule
and the second term, the enhancement term, is due to the momentum dependence of
the Hamiltonian.
The contribution of the momentum dependent part
is about 28\% of the total EWSR value in the oxygen isotopes considered here and
the term proportional to $C_{\alpha\beta\alpha'\beta'}$ describing
the effects of ground-state correlations is quite small (less than 1\%
of the total sum rule value).

In Figs.1$\sim$4, 
the $E1$ strength functions of $^{18\sim24}$O 
calculated in TDDM (solid line) are compared with those in 
TDHF (dotted line). The TDHF calculations presented here 
are equivalent to
the RPA calculations without any truncation of unoccupied single-particles states
because the TDHF equation for the
boosted single-particle wavefunctions $\psi_{\alpha}$ is solved in coordinate space.
The boundary condition for the continuum states, however, is not properly taken into
account in our calculation because all the single-particle wave functions are confined to 
the cylinder.
Therefore, the calculated strength functions slightly depend on the cylinder size.
The difference between the TDDM and TDHF
calculations is due to the effect of two-body correlations,
which may have two aspects: one is to induce ground-state
correlations which increase 
the occupation probabilities of weakly-bound neutron orbits.
The increase in low-lying strength ($E<15$MeV) seen in the
TDDM results for 
all the isotopes
is due to partial occupation of 
the neutron $2s_{1/2}$ and $1d_{3/2}$ states.
The comparison between the TDDM result for 
$^{24}$O and those for other isotopes suggests that the occupation of the neutron $2s_{1/2}$
state is responsible for the increase in the $E1$
strength in the very low energy region (around $E=5$MeV). This is because the appearance of 
a prominent peak at 6MeV in $^{24}$O is related to
nearly full occupation of the neutron $2s_{1/2}$
state. The other aspect of two-body correlations is to
increase the width of the giant dipole resonance (GDR) which is located
around 25MeV.  
There are background 2 particle - 2 hole states
which consist of 
$1\hbar\omega$
excitations of protons and $0\hbar\omega$ 
excitations of neutrons. The coupling of the GDR to these states leads to
the spreading of the GDR strength.
The fractions of the EWSR values depleted over the energy
range between 0MeV to 40MeV are about 85\% in TDDM and about 90\% in TDHF.
The EWSR values depleted below 15MeV
are shown in Fig.5. The ordinate indicates
the ratio of the EWSR value to the TRK
value.
The results in TDDM with $v_0=-230$MeVfm$^3$ (rhombus) are smaller
than the experimental data
except for $^{22}$O. In TDDM, the EWSR values in the low-energy
region crucially 
depend on the strength of the residual interaction.
Since the strength of the residual interaction is not well-known in
unstable nuclei, we performed TDDM calculations using slightly stronger
residual interaction with $v_0=-330$MeVfm$^3$
to increase $E1$ strength in the low-energy region.
The value of $v_0=-330$MeVfm$^3$ was used 
in our previous studies \cite{Toh1,Toh5} and found to give reasonable spreading widths
for giant resonances in stable nuclei \cite{Toh5}. 
As an example of the TDDM calculations with
$v_0=-330$MeVfm$^3$, we show the result for $^{20}$O in Fig.6. 
The increase in the strength below 10MeV is due to the increase
in the occupation probability of the neutron $2s_{1/2}$ 
state and the enhancement
of the peak at 15MeV is caused by further fragmentation of the GDR strength.
The EWSR values depleted below 15MeV are shown by circles 
in Fig.5. The magnitude of the low-lying strength is
increased and now becomes close to the experimental data
except for $^{22}$O.
The fact that the low-lying $E1$ strength in $^{20}$O 
is most increased with the use of the stronger residual interaction
may be explained by the large fragmentation of the
GDR strength as seen in Fig.6.
The TDDM results for $^{24}$O shown in Fig.5 significantly 
differ from the shell-model calculation \cite{Sag}.
The difference originates in the fact that
the shell-model calculation gives much smaller
spreading width to the GDR in $^{24}$O
than the TDDM and TDHF calculations do.
Since the large amount of the $E1$ strength is concentrated in the GDR,
the low-lying $E1$
strength calculated in the shell model becomes small in $^{24}$O.

Recently, surveying observed neutron separation energies
and interaction cross sections, A. Ozawa et al.\cite{Oza} has pointed out 
that
neutron number $N=16$ becomes a new magic number for nuclei near the
neutron drip line.
Shell model calculations done by 
Y. Utsuno et al.\cite{Utsu} have also shown
that a large energy gap between the $2s_{1/2}$ and $1d_{3/2}$ 
orbits makes
$N=16$ a new magic number for neutron-drip line nuclei.
They predict that the excitation energy of the first $2^+$ state in $^{24}$O
becomes the largest among the unstable oxygen isotopes as a 
consequence of the large energy gap \cite{Utsu}. 
Since the lowest-lying $E1$ mode
in $^{24}$O probably originates in the transition of a neutron from 
the $2s_{1/2}$ 
to the $2p_{3/2}$, its position and strength 
may also be used to explore the shell structure of $^{24}$O.
However, a quantitative discussion on 
the magicity of $N=16$ is difficult for
the present TDDM calculations. This is because the energy gap between 
the neutron $2s_{1/2}$ and $1d_{3/2}$ orbits does not significantly 
depend on $N$ in our calculations. In order to explore the magicity of
$N=16$ and its relation to the low-lying $E1$ modes, we might need larger
single-particle space and more realistic residual interactions.

In summary,
the strength functions of the isovector dipole resonances
in the even oxygen isotopes $^{18\sim24}$O were calculated by
using an extended version of TDHF known as TDDM. 
These calculations are much more improved and quantitative
than the previous ones \cite{Toh1} in two aspects: the inclusion of
spin-orbit force in a mean-field potential and the use of a correlated
ground state. The results were compared with recent
experimental data and also with a shell-model calculation. 
It was found
that if the strength of the residual interaction is appropriately chosen,
TDDM approximately reproduces the observed  
low-lying $E1$ strength.
It was also found that the TDDM prediction differs from
the shell-model calculation for $^{24}$O. It was discussed that this
originates in the difference in the spreading width of the GDR
between the two models.

\newpage
\begin{figure}[!t]
  \begin{center}
    \includegraphics[height=20pc]{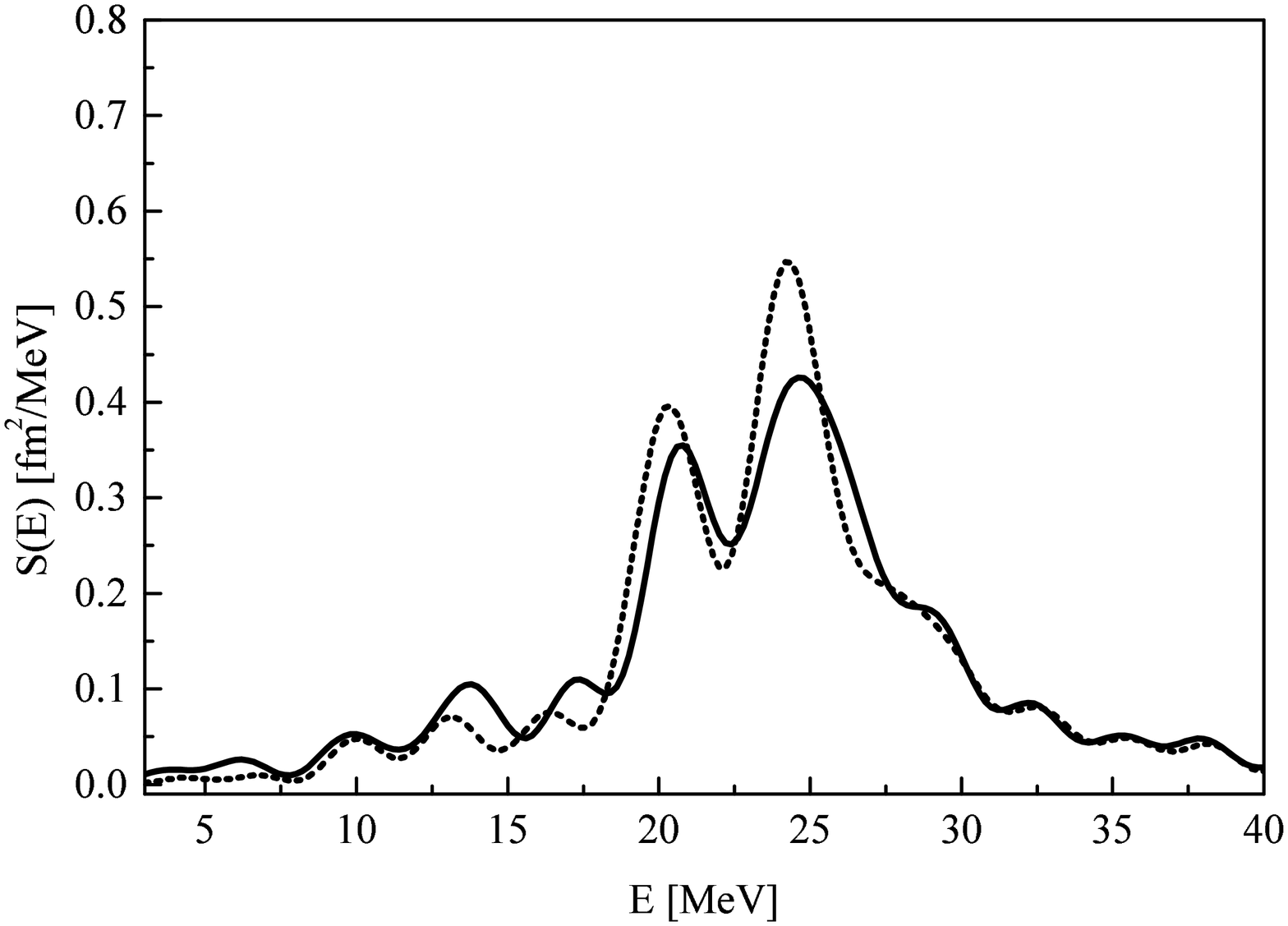}
  \end{center}
  \caption{Strength functions of isovector dipole resonances in $^{18}$O calculated 
in TDDM (solid line) and in TDHF (dotted line).}
\end{figure}

\vspace{1cm}
\begin{figure}[!t]
  \begin{center}
    \includegraphics[height=20pc]{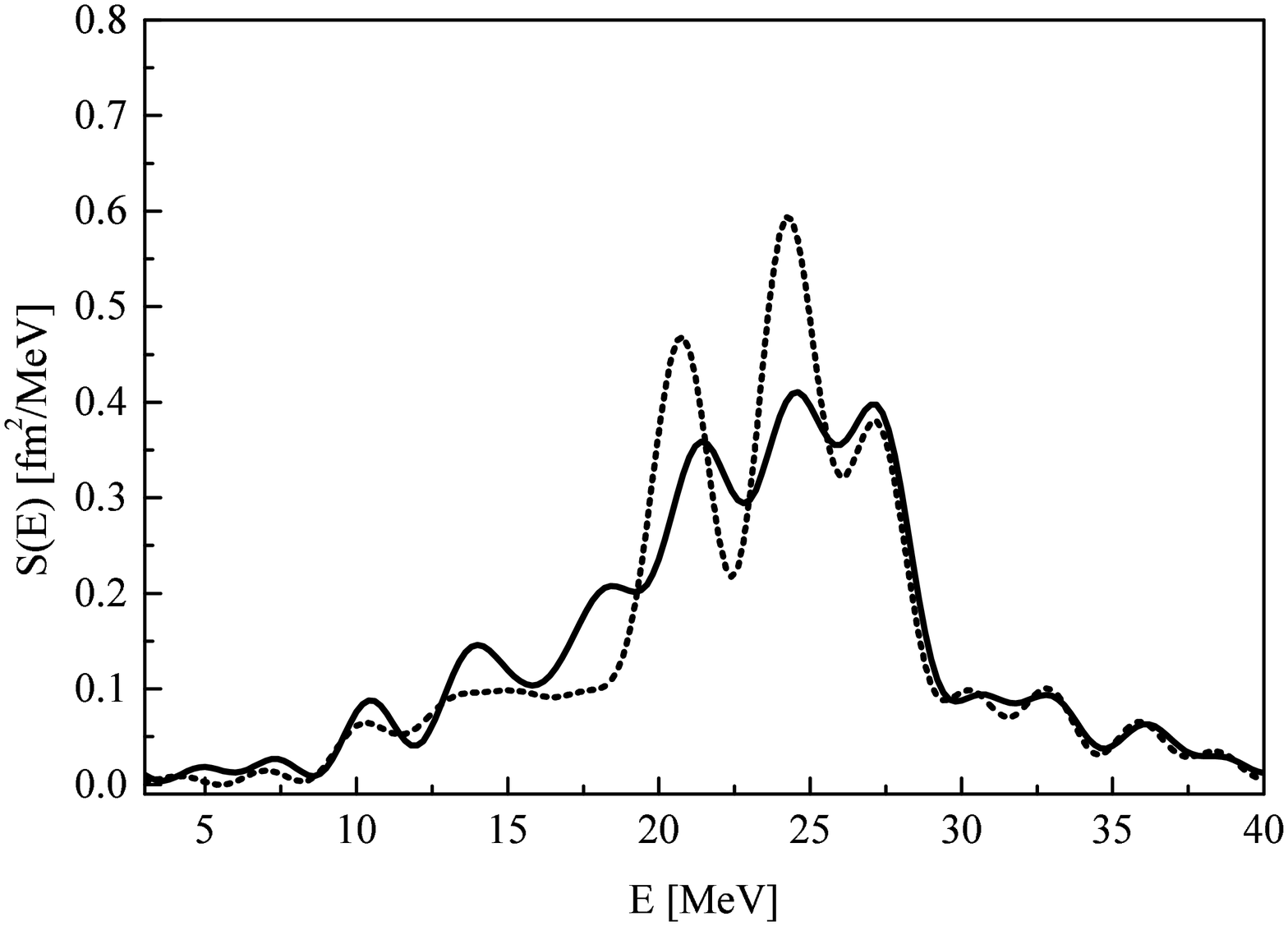}
  \end{center}
  \caption{Strength functions of isovector dipole resonances in $^{20}$O calculated 
in TDDM (solid line) and in TDHF (dotted line).}
\end{figure}

\newpage
\begin{figure}[!t]
  \begin{center}
    \includegraphics[height=20pc]{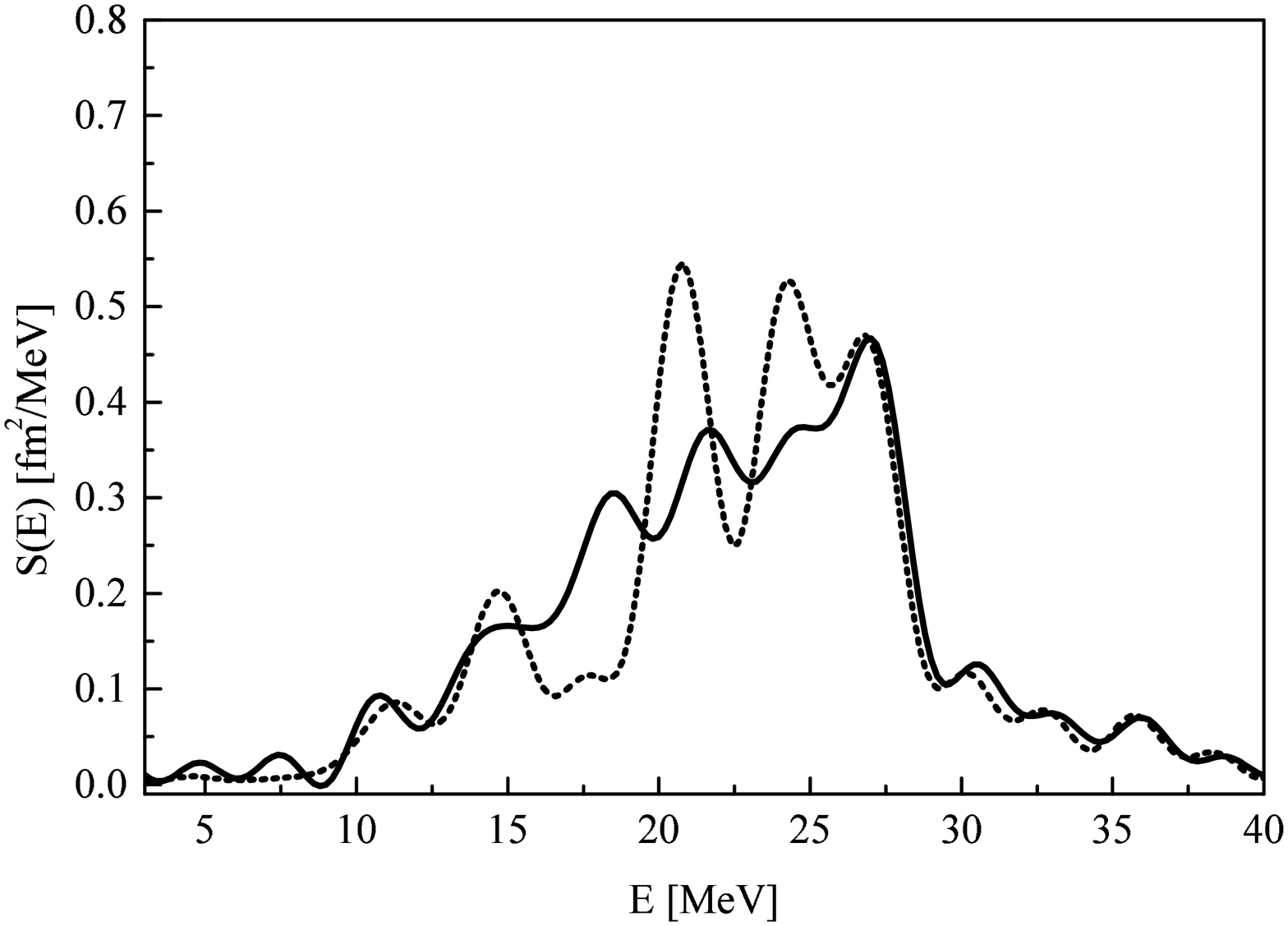}
  \end{center}
  \caption{Strength functions of isovector dipole resonances in $^{22}$O calculated 
in TDDM (solid line) and in TDHF (dotted line).}
\end{figure}

\vspace{1cm}
\begin{figure}[!t]
  \begin{center}
    \includegraphics[height=20pc]{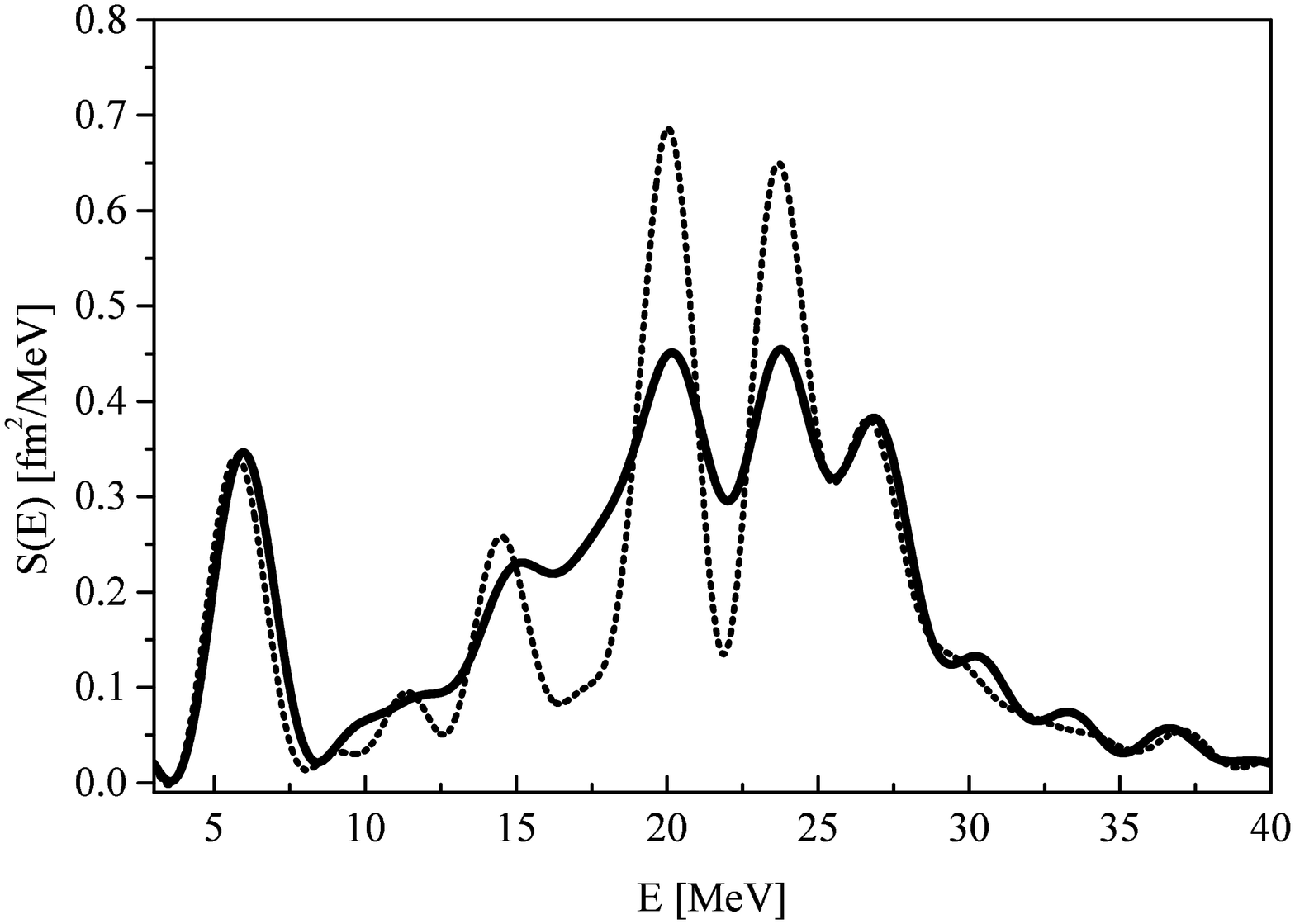}
  \end{center}
  \caption{Strength functions of isovector dipole resonances in $^{24}$O calculated 
in TDDM (solid line) and in TDHF (dotted line).}
\end{figure}
\newpage
\begin{figure}[!t]%4
  \begin{center}
    \includegraphics[height=20pc]{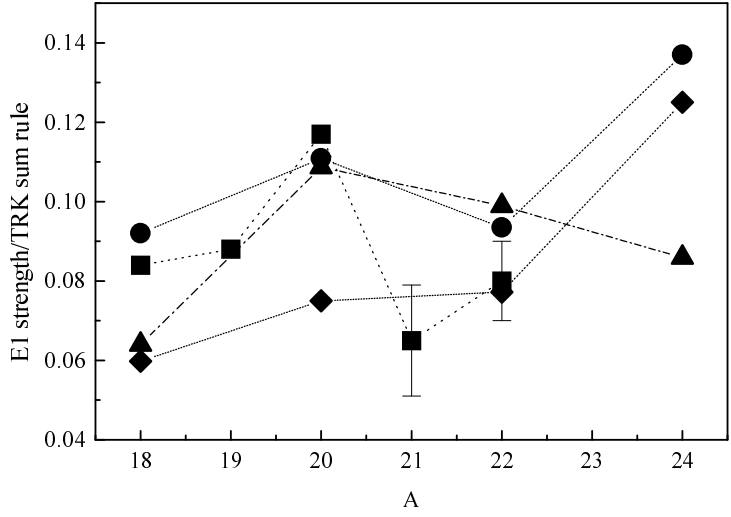}
  \end{center}
  \caption{Isotope dependence of the low-lying $E1$ strength. 
Square, triangle, rhombus, and circle indicate
the experiment$^{2)}$, the shell-model
calculation$^{3)}$, and the TDDM results with $v_0=-230$MeVfm$^3$
and $-330$MeVfm$^3$, respectively.}
\end{figure}

\vspace{1cm}
\begin{figure}[!t]
  \begin{center}
    \includegraphics[height = 20pc]{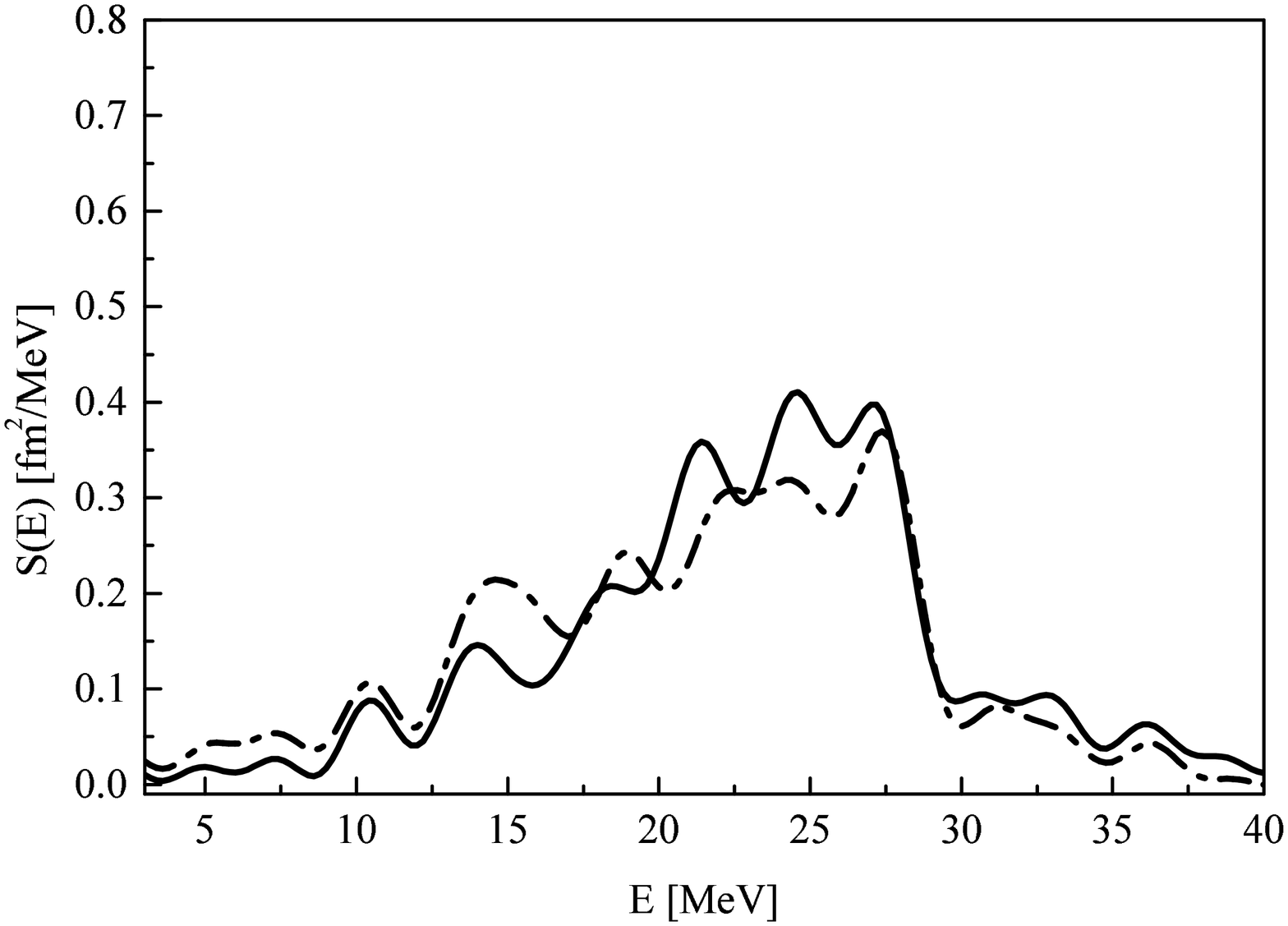}
  \end{center}
  \caption{Strength functions of isovector dipole resonances 
in $^{20}$O calculated with $v_0=-230$MeVfm$^3$ (solid line)
and with $v_0=-330$MeVfm$^3$ (dot-dashed line).}
\end{figure}

\end{document}